\documentclass{optica-article}

\journal{opticajournal} 

\articletype{Research Article}

\usepackage{lineno}

\begin{document}

\title{Event-based Asynchronous HDR Imaging by Temporal Incident Light Modulation}

\author{Yuliang Wu,\authormark{1} Ganchao Tan,\authormark{1} Jinze Chen,\authormark{1} Wei Zhai,\authormark{1*}  Yang Cao,\authormark{1,2} Zheng-Jun Zha,\authormark{1}}

\address{\authormark{1}Department of Automation, University of Science and Technology of China, Hefei, 230036, China\\
\authormark{2}Institute of Artificial Intelligence, Hefei Comprehensive National Science Center, Hefei, 230036, China\\
}

\email{\authormark{*}wzhai056@ustc.edu.cn}

\begin{abstract*} 
Dynamic Range (DR) is a pivotal characteristic of imaging systems. Current frame-based cameras struggle to achieve high dynamic range imaging due to the conflict between globally uniform exposure and spatially variant scene illumination. In this paper, we propose AsynHDR, a Pixel-Asynchronous HDR imaging system, based on key insights into the challenges in HDR imaging and the unique event-generating mechanism of Dynamic Vision Sensors (DVS). Our proposed AsynHDR system integrates the DVS with a set of LCD panels. The LCD panels modulate the irradiance incident upon the DVS by altering their transparency, thereby triggering the pixel-independent event streams. The HDR image is subsequently decoded from the event streams through our temporal-weighted algorithm. Experiments under standard test platform and several challenging scenes have verified the feasibility of the system in HDR imaging task.

\end{abstract*}

\section{Introduction}
An ideal imaging system is expected to efficaciously capture luminance and contrast information under various lighting conditions, encompassing a vast luminance range from approximately $10^{-2}$ lux in nocturnal starlight environments to $10^8$ lux in scenes illuminated by midday sunlight. In extremely low-light conditions, effective scene imaging can be achieved by enlarging the aperture size and prolonging exposure durations, while in brightly illuminated environments, sensor overexposure can be avoided by using a smaller aperture and a shorter exposure time. However, in high dynamic range scenarios, imaging systems that use globally uniform sampling, exposure, and light input control face challenges due to limitations in sampling bit depth and electron well capacity. During each imaging process, only a limited number of pixels on the sensor can achieve optimal exposure, while other pixels fail to perceive the scene accurately due to inappropriate illumination parameter settings.

Current mainstream methods for high dynamic range imaging can be divided into multi-exposure fusion (MEF) and spatial light modulators-based (SLMs-based) approaches. As the most widely applied HDR imaging technique, MEF entails capturing multiple frames with varying exposure parameters on CMOS/CCD sensors, followed by meticulous selection and fusion of regions with optimal exposure across the frames to generate an HDR image\cite{Debevec_Malik_2008, jinno2011multiple, kang2003high, mase2005wide, hasinoff2016burst, hasinoff2009multiple,martinez2015adaptive, tocci2011versatile, hajisharif2015adaptive}. However, the demand for repetitive sampling of frames poses several challenges to MEF. Single-sensor MEF methods \cite{Srikantha_Sidibé_2012,silk2012high,karadjuzovic2017assessment}  are troubled by the ghosting artifacts due to temporal misalignment. Multi-sensor methods \cite{yamashita2017hdr,seshadrinathan2017high,huynh2019high} face challenges such as sensor registration and structural complexity. And MEF methods obtain multiple images by reusing Bayer matrices at the expense of sacrificing sensor spatial resolution \cite{hajisharif2015adaptive, Nayar_Mitsunaga_2002}.
In contrast to the MEF approaches, SLMs-based methods involve modulating the irradiance incident upon the sensor pixel-by-pixel. By utilizing SLMs\cite{saxena2015structured} such as DMD \cite{adeyemi2009applications, riza2016demonstration, feng2016per, mazhar202096, guan2021pixel, nayar2006programmable, qiao2015design, zhou2019design,feng2017digital} or LCD \cite{mannami2007adaptive, nayar2003adaptive}, the incident light of each pixel is independently attenuated based on the intensity of incident light, ensuring that it falls within the effective working range of the sensor. However, the introduction of high-cost SLMs leads to a decrease in imaging quality, and the parameters of the SLMs in the structure are scene-dependent, as real-time feedback adjustments are required for different scenarios.

The advent of asynchronous sensors introduces the potential to develop HDR imaging systems with pixel-independent sampling.
By leveraging asynchronous sensors such as Dynamic Vision Sensors (DVS), imaging systems can break free from the constraints of globally uniform pixel sampling, constructing an HDR system with pixel-independent triggering. Previous work has extensively explored DVS-based image reconstruction, such as estimating the scene radiance by leveraging motion-triggered event streams  \cite{Rebecq_Ranftl_Koltun_Scaramuzza_2019,yang2023learning}, or constructing motion-independent DVS imaging systems using actively controlled light sources to modulate scene brightness \cite{han2023high, muglikar2021esl, takatani2021event,huang2021high,liu2023event,fu2023fast}. However, the preceding approaches are compromised due to limited scene information contained in the sparse events triggered by motion. And the systems incorporating active light sources are not applicable to HDR outdoor scenarios or HDR scenes containing light sources.
In contrast, our proposed method leverages the asynchronous sensing features of the dynamic vision sensor to achieve HDR imaging, enabling operation in various HDR scenarios. 

In this paper, we develop the \textbf{Asyn}chronous \textbf{HDR} imaging system (\textbf{AsynHDR}), which triggers event streams by introducing temporal variations in the system's incident light intensity. Compared to active light-triggered imaging systems \cite{takatani2021event,han2023high}, the AsynHDR system achieves HDR scene imaging ability by combining the proportional attenuation of incident light with DVS's independent pixel triggering mechanism.
The optical architecture of the AsynHDR system consists of a DVS, two LCD panels, a beam splitter, and a signal generator. The LCD panels dynamically modulate the transmittance to control the incident light in the system. The DVS in the system triggers event streams on a per-pixel basis within suitable exposure ranges.
Building upon the hardware system, we further propose a temporal-weighted algorithm to replace the direct integration method for the reconstruction of scene radiance from event streams. Combined with subsequent threshold correction processing, it significantly enhances imaging signal-to-noise ratio (SNR) and quality.

Our contributions can be summarized as follows:
\begin{itemize}
    \item First, we discern the efficacy of sensor pixels operating independently in tackling HDR challenges. Combining this observation with the operating principles of DVS, we propose the construction methodology for DVS-based HDR imaging systems.
    \item Second, by modulating the incident light using LCD panels, the AsynHDR system constructed by us can recover scene radiance from the triggered event stream, and we propose a temporal-weighted method to enhance imaging quality.
    \item Third, the experiments under the challenging light-source included and outdoor HDR scenarios validate the system's high-quality HDR imaging capability, and confirm the viability of DVS conducting passive imaging without the aid of frame-based cameras or active light sources.
\end{itemize}

\section{Principle and Method}
In the following three subsections, we will introduce the principles for constructing an asynchronous HDR imaging system, the optical architecture of our asynchronous HDR imaging system, and the temporal-weighted algorithm for reconstructing HDR images from event streams.

\subsection{Methodology for AsynHDR Imaging System}
Constructing a pixel-independent HDR imaging system requires selecting an asynchronous sampling sensor as the sensing component. This paper outlines the construction methodology for an asynchronous HDR imaging system using DVS.
Unlike frame-based cameras, each pixel in the DVS array operates independently, triggering events based on the event-triggering mechanism. Each event is defined as:
\begin{equation}
    e_{i}:=\left(x_{i}, y_{i}, t_{i}, p_{i}\right),
   \label{eq1}
\end{equation}
where $(x_{i}, y_{i})$ represents the pixel coordinates, $t_{i}$ is the timestamp of the event, and $p_{i} \in\{-1,+1\}$ indicates the polarity of the event. An event is triggered when the change of logarithmic intensity of the pixel, $log\mathbf{I}(x, y, t):=\overline{\mathbf{I}}(x,y,t)$, exceeds the triggering threshold $c$:
\begin{equation}
   \left|\Delta \overline{\mathbf{I}}(x_{i}, y_{i},t_{i})\right|=\left|\overline{\mathbf{I}}\left(x_{i}, y_{i}, t_{i}\right)-\overline{\mathbf{I}}\left(x_{i}, y_{i},t_{i}-\Delta t_{i}\right)\right| \geq c,
   \label{eq2}
\end{equation}
where $\Delta t$ is the time interval from the previous event to the current event at the position $(x_i, y_i)$.

To construct an HDR imaging system based on DVS, it is essential to obtain sufficiently informative event streams. In addition to utilizing changes in scene illumination and object motion to generate events, we can also dynamically alter the camera's incident light to trigger DVS event streams by incorporating devices such as optical valves into the optical path of the system. The incident light at the sensor pixel $(x, y)$ can be modeled as follows:
\begin{equation}
   \mathbf I(x,y,t) = f(t)\mathbf L(x,y,t),
   \label{eq3}
\end{equation}
where $f(t)$ is the temporal modulation factor for the imaging system's incident light, and $L$ is the scene radiance component incident on the pixel.

Assuming nearly constant scene radiance over a short period, the event-triggering mode is as follows:
\begin{equation}
\begin{split}
\left|\Delta \overline{\mathbf{I}}\left(x,y,t\right)\right| = 
&\left| \log f(t)\mathbf L(x, y, t) -  \log f(t-\Delta t)\mathbf L(x, y, t-\Delta t) \right|\\
&=\left| \log f(t)\mathbf L(x, y) -  \log f(t-\Delta t)\mathbf L(x, y) \right|\\
&=\left|\log f(t)-\log f(t-\Delta t)\right|\geq c.
\end{split}
\label{eq4}
\end{equation}
We can observe that the logarithmic threshold event triggering characteristics of DVS, coupled with the separable form of the incident light modulation function, result in the triggering timestamps of all pixels being solely dependent on the light modulation function $f(t)$. These timestamps are independent of the scene light intensity component $L$. Uniformly adjusting the incident light of the system triggers events with consistent timestamps and does not contain any scene radiance information.

Therefore, to encode scene radiance information into the time stamps of event streams, the temporal variation component in Eq.~\ref{eq3} needs to be designed in a form inseparable from the scene radiance $\mathbf{L}$, 
\begin{equation}
 \begin{aligned}  
  \mathbf I(x,y,t) = f_{inseparable}(t,\mathbf L(x,y)).
 \end{aligned} 
\label{eq5}
\end{equation}
Such as 
\begin{equation}
 \begin{aligned}  
  \mathbf I(x,y,t) = f(t)\mathbf L(x,y) + g(t),
 \end{aligned} 
\end{equation}
of our system, under this setting, the event triggering mode is as follows:
\begin{equation}
\begin{split}
\left|\Delta \overline{\mathbf{I}}\left(x,y,t\right)\right|
&=\left| \log [f(t)\mathbf L(x, y)+g(t)] -  \log [f(t-\Delta t)\mathbf L(x, y)+g(t)] \right|\geq c.
\end{split}
\label{eq6}
\end{equation}
In this modulation, the scene component won't be eliminated as in Eq.~\ref{eq4}. The information about scene radiance $\mathbf L$ can be encoded into the temporal characteristics of the event streams.
\subsection{Construction of AsyHDR Imaging System}
\begin{figure*}[t]
  \centering
  \includegraphics[width=0.8\textwidth]{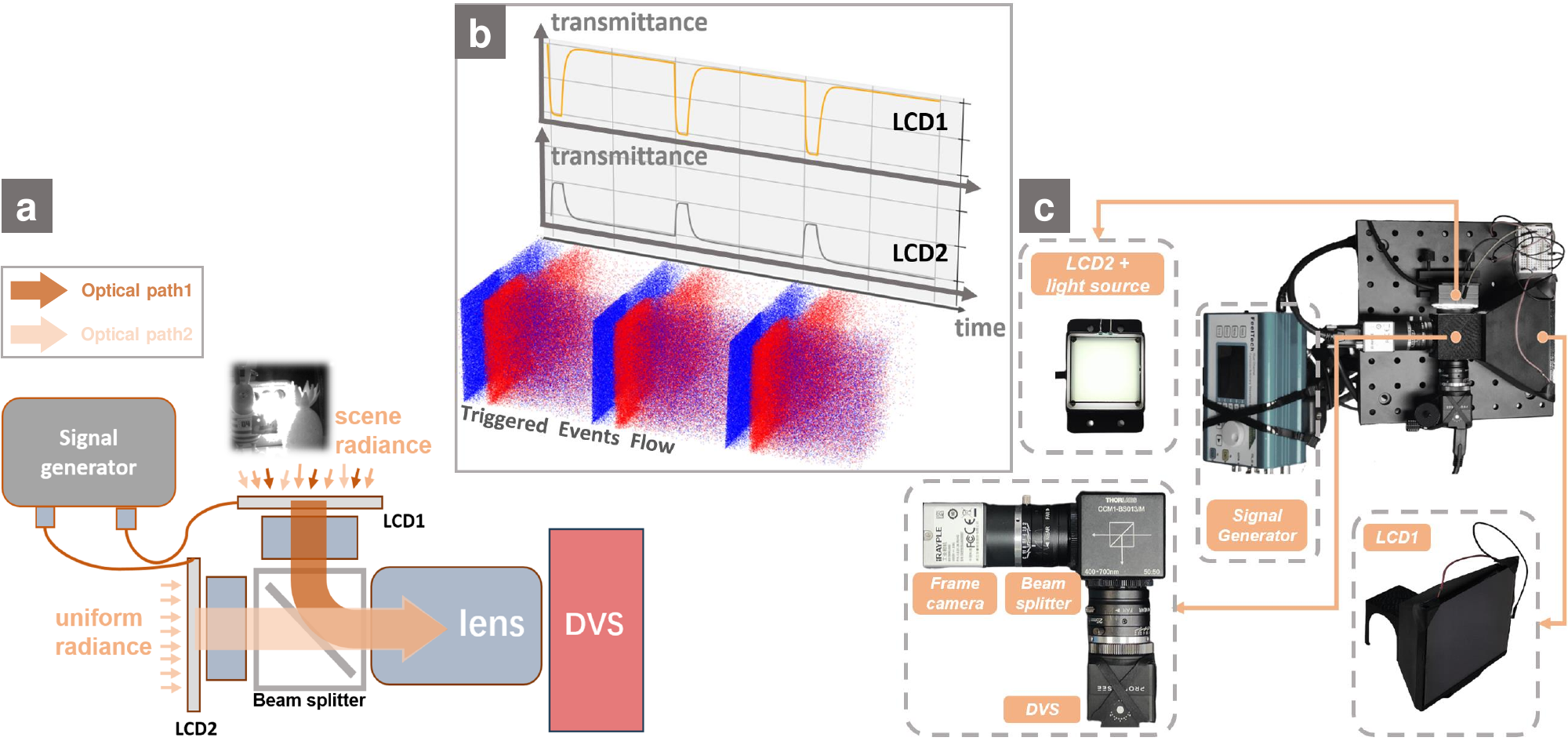}
  \caption{
  (a) Optical schematic diagram of our AsynHDR system. (b) Event triggering demonstration. The point cloud diagram illustrates events triggered by the dynamic modulation of LCD panels, where red represents positive events and blue represents negative ones. (c) Physical demonstration of the system.}
  \label{fig: system architecture}
\end{figure*}

With the theoretical foundation from the previous subsection, we constructed an AsynHDR system where the incident light triggering events are modulated by LCD panels, as shown in Fig.~\ref{fig: system architecture}.
The system consists of a DVS, LCD panels, a signal generator, beam splitter, and lenses. The sensor irradiance ($\mathbf{I}$) incident on the DVS pixels array is obtained by proportionally attenuating the environmental light scene radiance ($\mathbf{L}$) through the LCD panels in the optical path and then transmitting through the lenses. The mathematical expression for this process is:
\begin{equation}
   \mathbf I(x,y,t) = T(t)k_{lens}\mathbf L(x,y,t).
   \label{eq7}
\end{equation}
Here, $T(t)\geq 0$ represents the transmittance of the LCD panels, $k_{lens}$ is the attenuation coefficient of the lens, and $\mathbf I(x,y,t)$ represents the irradiance component projected onto the DVS sensor pixel $(x,y)$ at time $t$ in this optical path.

In the AsynHDR system, the irradiance projected onto the sensor is composed of two beams modulated by LCD panels:
\begin{equation}
      \mathbf{I}\left(x, y,t\right)=\mathbf{I}_{1}(x,y,t)+\mathbf{I}_{2}(t),
   \label{eq8}
\end{equation}
$I_{1}(t)$ represents the sensor irradiance component of the scene incident light, while $I_{2}(t)$ is a uniformly weak incident light component that varies only with time which is used to provide a consistent starting sampling value for all pixels,
\begin{equation}
\left\{  
 \begin{aligned}  
 \mathbf{I}_{1}(x,y,t) &= T_{1}(t)k_{lens}\mathbf L(x, y),t_0<t<t_1   \\  
 \mathbf{I}_{2}(t) &= T_{2}(t)k_{lens}\mathbf L_{const},t_0<t<t_1\\   
 \end{aligned} 
\right..
   \label{eq9}
\end{equation}
The critical formula for system event triggering is as follows:
\begin{equation}
\begin{split}
\left|\Delta \overline{\mathbf{I}}\left(x,y,t\right)\right| = 
&\left| \log[T_{1}(t)k_{lens}\mathbf L(x, y) + T_{2}(t)k_{lens}\mathbf L_{const}] - \right. \\
&\left. \log[T_{1}(t-\Delta t)k_{lens}\mathbf L(x, y) + T_{2}(t-\Delta t)k_{lens}\mathbf L_{const}] \right| \geq c.
\end{split}
   \label{eq10}
\end{equation}
The information of the scene radiance component $\mathbf L(x, y)$ corresponding to pixel point $(x, y)$ is encoded in the event stream, and HDR image reconstruction can be achieved through appropriate processing.

\subsection{Reconstruction of HDR Intensity Images from the Event Streams}
Previous approaches have restored scene radiance by directly integrating events \cite{muglikar2021esl}.
However, this method results in very few gray levels and is severely degraded by noise. We incorporate temporal information into the reconstruction process to achieve a low noise level and a nuanced gray-scale response.

Let's consider any pixel points $\left(x_{1}, y_{1}\right)$ and $\left(x_{2}, y_{2}\right)$. On the first optical path, the LCD transmission function $T_{1}(t)$ is monotonically increasing. On the second optical path, the LCD transmission function $T_{2}(t)$ is monotonically decreasing, and $|\frac{T_{2}(t)}{dt}|<|\frac{T_{1}(t)}{dt}|, \forall t\geq 0$.
Assuming both pixel points can trigger more than $k$ events, the critical condition equation for event triggering is:
\begin{equation}
\begin{split}
\log[T_{1}(t^{k}_{1})\mathbf L\left(x_{1}, y_{1}\right) + T_{2}(t^{k}_{1})\mathbf{L}_{const}] - \log[T_{1}(0)\mathbf L \left(x_{1},y_{1}\right) + T_{2}(0)\mathbf{L}_{const}] = \\
\log[T_{1}(t^{k}_{2})\mathbf{L} \left(x_{2},y_{2}\right) + T_{2}(t^{k}_{2})\mathbf{L}_{const}] - \log[T_{1}(0)\mathbf{L}\left(x_{2},y_{2}\right) + T_{2}(0)\mathbf{L}_{const}]=kc,
\end{split}
\label{eq11}
\end{equation}
where $kc$ represents the $k$-th order event triggering threshold.
Substituting $T_{2}(0)=1,T_{1}(0)=0$ into the above equation yields:
\begin{equation}
\begin{split}
T_{1}(t^{k}_{1})\mathbf{L}\left(x_{1}, y_{1}\right) + T_{2}(t^{k}_{1})\mathbf{L}_{const}= 
T_{1}(t^{k}_{2})\mathbf{L}\left(x_{2},y_{2}\right) + T_{2}(t^{k}_{2})\mathbf{L}_{const}
=\exp(kc+\log \mathbf{L}_{const}).
\end{split}
\label{eq12}
\end{equation}
We can deduce that the relationship between the triggering moments $t^{k}_{1},t^{k}_{2}$ and the scene illumination $L\left(x_{1}, y_{1}\right),L\left(x_{2}, y_{2}\right)$:
\begin{equation}
\begin{split}
 \begin{aligned}  
 &t^{k}_{1}<t^{k}_{2}, \\
 s.t.\quad &\mathbf L\left(x_{1}, y_{1}\right)\textgreater\mathbf L\left(x_{2}, y_{2}\right)\geq\mathbf{L}_{const}   .\\  
 \end{aligned} 
\end{split}
\label{eq:temporal relationship}
\end{equation}

\begin{figure*}[t]
  \centering
  \includegraphics[width=0.8\textwidth]{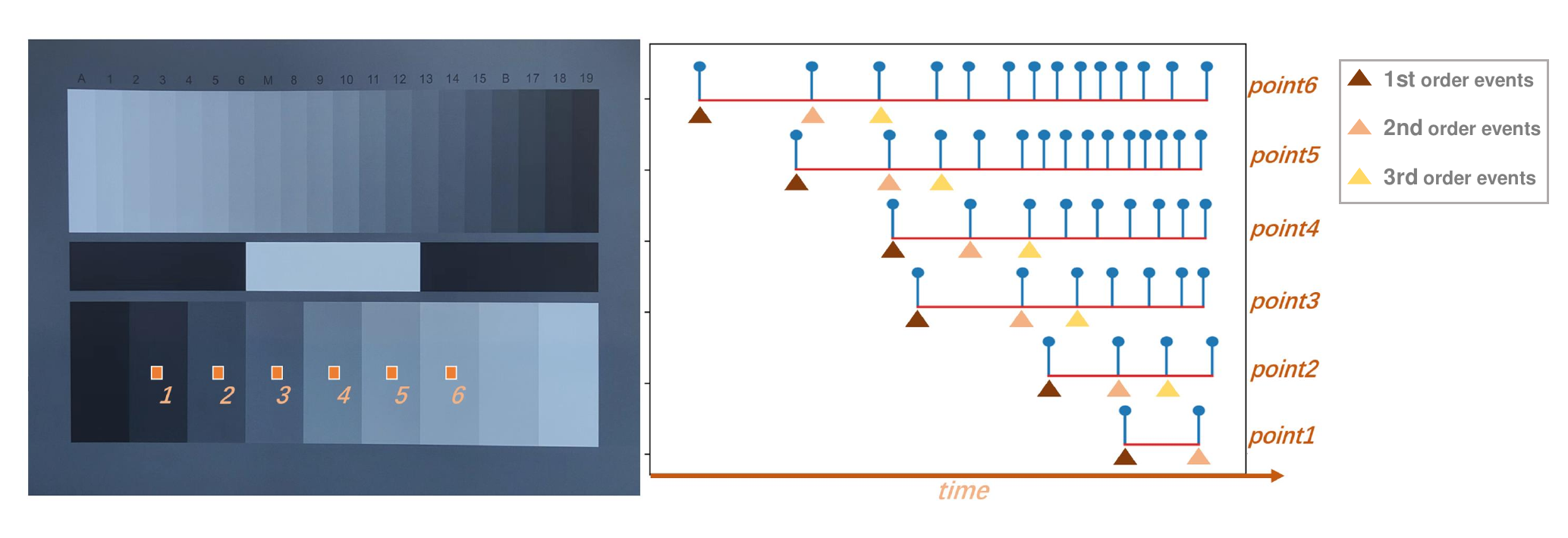}
  \caption{Pixel-wise presentation of events triggered at different light intensities.
  (a) Sampling points chosen from continuous stepped radiance levels on a gray-scale test card. (b) The events triggered at different points along the timeline, where blue lines represent the positive events ($p_{i}$=+1), and colored triangles indicate different order events for each pixel.
  }
  \label{fig:pointtrigger}
\end{figure*}

This implies that the relative brightness information of pixels is encoded in the temporal information of event triggering. As shown in Fig.~\ref{fig:pointtrigger}, we utilized the AsynHDR system to capture the gray-scale gradient test card, showcasing event streams recorded in different scene radiance regions. Specifically, brighter pixels reach the triggering threshold earlier, resulting in smaller event timestamps.

Based on the above conclusions, we designed a temporal-weighted algorithm to extract information from the event stream and map it into an intensity image:
\begin{equation}
\begin{split}
img(x,y) = \mathbf{L}_{const}*exp(\sum_{\substack{x_i=x,y_i=y}}f(t_i)*p_i*c),
\end{split}
\label{eq14}
\end{equation}
where $img(x, y)$ represents the intensity of the pixel at coordinates $(x, y)$ in the recovered image. $f(t)$ is the temporal-weighted function, assigning weights to each event based on their triggering timestamps.
Considering $\mathbf{L}\left(x_{1}, y_{1}\right)\textgreater\mathbf{L}\left(x_{2}, y_{2}\right)\geq\mathbf{L}_{const}$, we aim to ensure the monotonicity of the system, meaning that the pixel intensities $img\left(x_{1},y_{1}\right)$ and $img\left(x_{2}, y_{2}\right)$ recovered from the corresponding pixel event streams maintain the same size relationship as $\mathbf{L}\left(x_{1}, y_{1}\right)$ and $\mathbf{L}\left(x_{2}, y_{2}\right)$.
Combining the previous derivation, the weighting function $f(t)$ only needs to decrease monotonically in the time domain to ensure a monotonically consistent system. Subsequently, we will demonstrate how the introduction of the function $f(t)$ enhances imaging quality.

 In our DVS-HDR imaging system, noise mainly originates from two aspects: The pseudo-events triggered by fluctuations in sensor dark current, and the inconsistency in event thresholds among pixels\cite{wang2020event}. We mitigate the impact of pseudo-events on imaging by introducing and optimizing $f(t)$. Simultaneously, we estimate an event threshold correction map to eliminate the multiplicative fixed pattern noise (FPN) caused by threshold inconsistency, further enhancing the imaging signal-to-noise ratio (SNR).

We define pseudo-events and valid-events as $e^{ps}_{i}$ and $e^{val}_{i}$.
Substituting into the reconstruction formula, the value of the reconstruction image at position $(x, y)$ can be expressed as follows:
\begin{equation}
\begin{split}
img(x,y) &= \mathbf{L}_{const}*exp[(\sum_{\substack{x^{val}_i=x,y^{val}_i=y}}f(t^{val}_{i})*p_i^{val}+\sum_{\substack{x^{ps}_{i}=x,y^{ps}_{i}=y}}f(t^{ps}_{i})*p^{ps}_{i})*c]  
\\&=\mathbf{L}_{const}*(E_{val}(x,y)*E_{pseudo}(x,y)).
\end{split}
\end{equation}
Assuming $\mathbf{L}\left(x_{1}, y_{1}\right)\textgreater\mathbf{L}\left(x_{2}\right)$, we define the difference term as:
\begin{equation}
\begin{split}
 \delta(x_1,y_1,x_2,y_2) =& \frac{img(x_1,y_1)}{img(x_2,y_2)}=\frac{E_{val}(x_1,y_1)}{E_{val}(x_2,y_2)}*\frac{E_{pseudo}(x_1,y_1)}{E_{pseudo}(x_2,y_2)}\\
  =& \delta_{val}(x_1,y_1,x_2,y_2) * \delta_{ps}(x_1,y_1,x_2,y_2).
\label{eq:difference term}
\end{split}
\end{equation}
We aim to identify a method that amplifies $\delta_{val}$ while slightly affecting $\delta_{ps}$ to enhance the imaging quality.
According to the mechanism of Eq.~\ref{eq:temporal relationship}, the reconstruction intensity of pixel of the same-level valid events with higher intensity triggers earlier than those with lower intensity,  while pseudo events do not possess this characteristic due to equi-probable triggering in the time domain. 
The relationship between $\delta_{val}$ with and without the weighting function $f(t)$ can be expressed as:
\begin{equation}
\frac{E_{val}(x_1,y_1)}{E_{val}(x_2,y_2)}=\frac{\sum_{\substack{x_i=x_1,y_i=y_1}}f(t_i)*p_i}{\sum_{\substack{x_i=x_2,y_i=y_2}}f(t_i)*p_i}\textgreater
\frac{\sum_{\substack{x_i=x_1,y_i=y_1}}p_i}{\sum_{\substack{x_i=x_2,y_i=y_2}}p_i}
\label{eq19}.
\end{equation}
Considering $\frac{f(t)}{dt}<0$, Eq.~\ref{eq19} can be proven, and $\delta_{val}$ achieves amplification.
The most straightforward monotonically decreasing function takes on a linear form, denoted as  $f(t) = 1-t, (0<t<1)$. And in the experiments section, we analyze the enhancement effects of different temporal weighting approach by standard test, and replaced linear funcion with exponential funtion $f(t) = 2^{-k * t}, \quad k \in \mathbb{Z}^{+},(0 < t < 1)$ to further amplifying the SNR of the results. 
The final reconstruction formula is as follows:
\begin{equation}
\begin{split}
img(x,y) = \mathbf{L}_{const}*exp(\sum_{\substack{x_i=x,y_i=y}}2^{-k*t}*p_i*c).
\end{split}
\label{eq20}
\end{equation}

To address the noise introduced by varying pixel event triggering thresholds, we introduced a calibration step to correct the imaging system, reducing fixed pattern noise (FPN) caused by inconsistent pixel parameters. 
We acquired the correction tensor (c-map) through the iterative acquisition of images from a uniformly illuminated light box, followed by the averaging of the obtained results.
The c-map was then used as correction parameters for image reconstruction (see results in Fig.~\ref{fig:weighting result}).

\section{Experiments}
\begin{figure*}[t]
    \centering
  \begin{minipage}[b]{0.8\textwidth}
    \centering
    \resizebox{1\textwidth}{!}
    {\includegraphics[width=\textwidth]{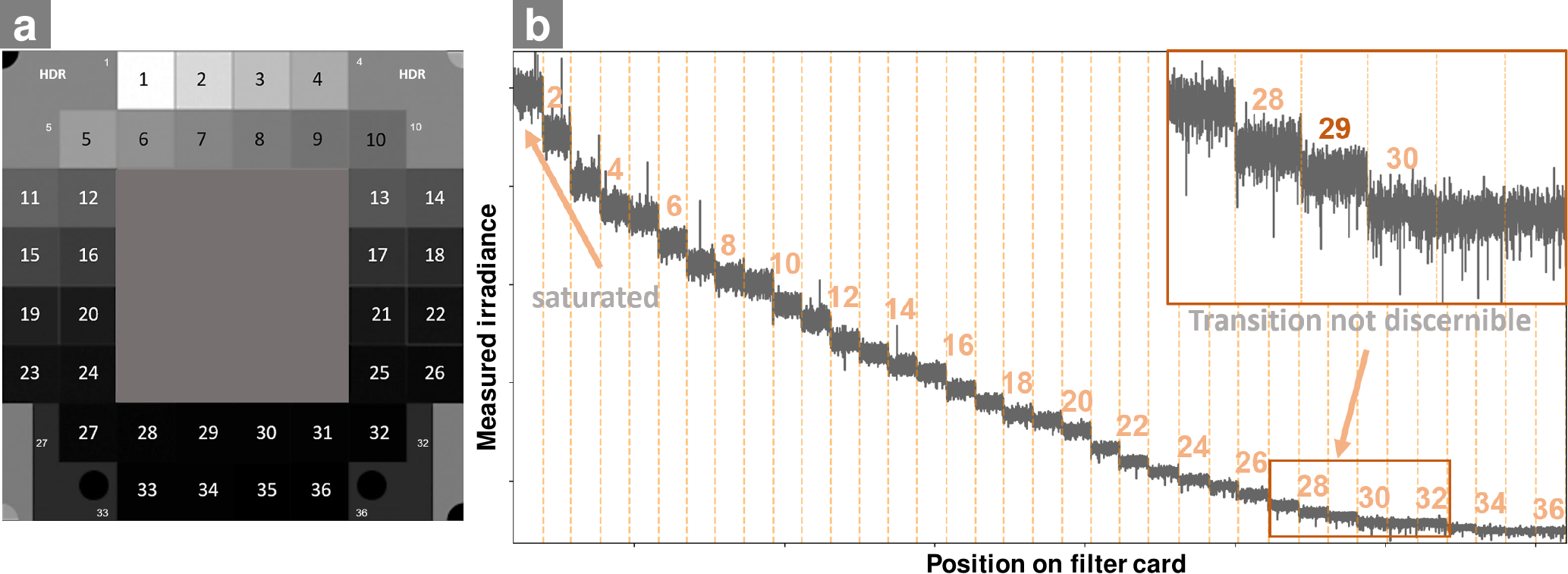}
    }
  \end{minipage}
  \begin{minipage}[t]{0.8\textwidth}
    \centering
    \resizebox{1\textwidth}{!}
    {\begin{tabular}{|c|c|c|c|c|c|c|c|c|c|c|c|c|c|c|c|c|c|c|}
        \hline
        \textbf{level} & \textbf{1} & \textbf{2} & \textbf{3} & \textbf{4} & \textbf{5} & \textbf{6} & \textbf{7} & \textbf{8} & \textbf{9} & \textbf{10} & \textbf{11} & \textbf{12} & \textbf{13} & \textbf{14} & \textbf{15} & \textbf{16} & \textbf{17} & \textbf{18} \\ \hline
        \textbf{density}&0.00  & 0.10 &  0.20 &  0.30   &0.40   &0.50 &1.20 &  1.30 &  1.50 &  0.60 &  0.70 &  0.80  & 0.90 & 1.00&1.10 &  1.70 &  1.90  & 2.10\\ \hline
        \textbf{level}& \textbf{19} & \textbf{20}& \textbf{21} & \textbf{22} & \textbf{23} & \textbf{24} & \textbf{25} & \textbf{26} & \textbf{27}& \textbf{28} & \textbf{29} & \textbf{30}& \textbf{31} & \textbf{32} & \textbf{33} & \textbf{34} & \textbf{35} & \textbf{36}   \\ \hline 
        \textbf{density}  &  2.30&2.50 &2.70& 2.90 & 3.43  & 3.73  & 4.02  & 4.32  & 4.63  & 4.92  & 5.23 &5.52&5.82&6.27& 6.72  & 7.17  & 7.63  & 8.22  \\ \hline
        \end{tabular}}
  \end{minipage}
\caption{Dynamic range test and denoise algorithm experiment.
  \\(a) The stepped transmission brightness test card. (b) Illustration of the dynamic range test curve for the system.
  The table at the bottom displays the transmittance density of different filters for the filter array.}
\label{fig:stepchart exper}
\end{figure*}
\begin{figure*}[t]
  \centering
  \includegraphics[width=0.8\textwidth]{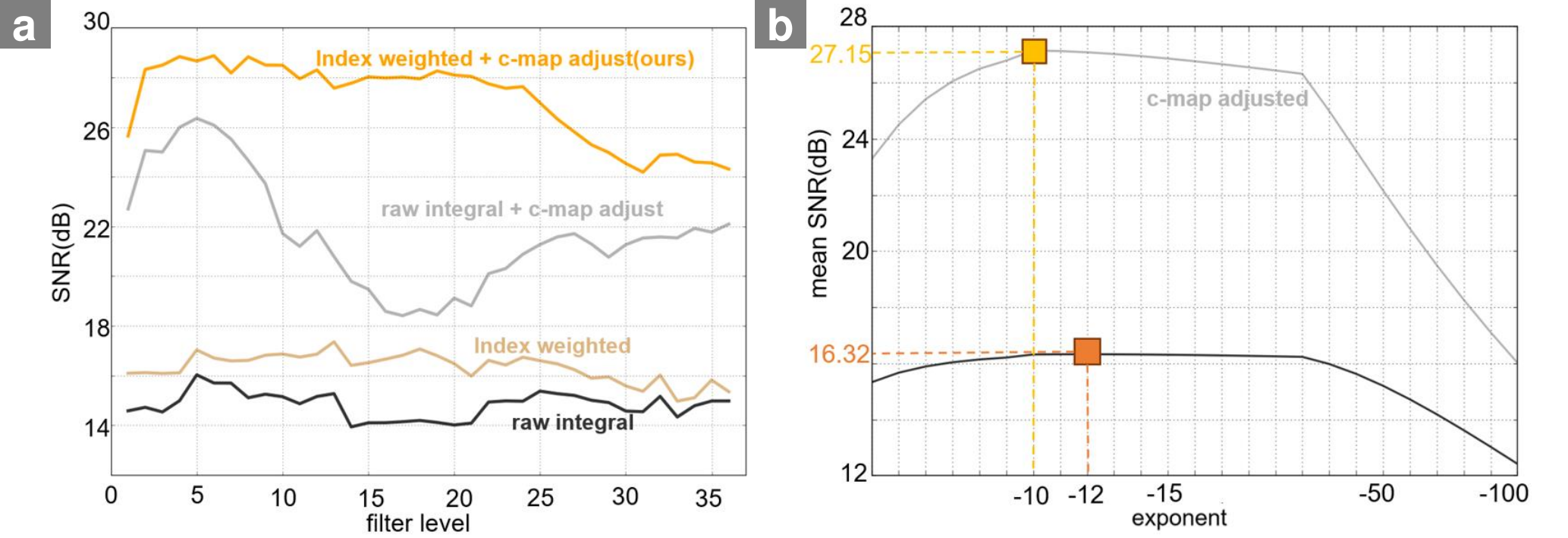}
\caption{Analysis of Algorithm SNR.
(a) The SNR curves of different uniform radiance regions on the test card under various temporal weighting enhancement strategies with/without c-map adjustment. (b) The average step-radiance SNR under different k-factor exponential temporal weighting.}
\label{fig:SNR curves}
\end{figure*}

We employ a standard testing platform to evaluate the dynamic range of the AsynHDR system and showcase the performance of the temporal-weighted algorithm.
The platform consists of a high-intensity uniform lightbox (160,000 lux illuminance) and a density filter array. By imaging a set of neutral density filters varying in transmittance on the array, we tested the dynamic range of the system. As shown in Fig.~\ref{fig:stepchart exper}a, the filter array exhibits uniformity in each region with varying transmittance density ($Dt$), which is computed as follows: 
\begin{equation*}
\begin{split}
 Dt = \lg \frac{P_{o}}{P_{t}},
\end{split}
\end{equation*}
where $P_{o}$ represents incident light and $P_{t}$ represents transmitted light.

The HDR test results are displayed in Fig.~\ref{fig:stepchart exper}b, revealing that the AsynHDR system exhibits perceptual sensitivity to brightness variations across filter levels 2 through 29. The values of $Dt$ for the 2nd-order filter ($Dt_{0}$) and the 28th-order filter ($Dt_{1}$) are 0.1 and 5.23.
The dynamic range of the AsynHDR system is calculated as follows:
\begin{equation*}
\begin{split}
 DR = 20\lg \frac{L_{max}}{L_{min}} = 20(Dt_{0}-Dt_{1}) =102.6dB.
\end{split}
\end{equation*}

We conducted experiments on the same testing platform to validate the denoising capability of the temporal-weighted algorithm.
As shown in Fig.~\ref{fig:SNR curves}(a), by calculating the Signal-to-Noise Ratio (SNR) in different filters of the array, we demonstrate the denoising ability of various event processing methods at different brightness levels.
The SNR is calculated as follows:
\begin{equation*}
\begin{split}
 SNR =10 \lg \left(\frac{\mu^{2}}{\sigma^{2}}\right),
\end{split}
\end{equation*}
where $\mu$ represents the average value of pixels irradiance,
and $\sigma$ represents the standard deviation of the noise.
In the event encoding step described in the previous section, we employ an weighting method to suppress noise. 
Considering the reconstructed image combines information from both event timestamps and the accumulated number of events, increasing the value of $k$ in the weighting method indefinitely doesn't guarantee improved reconstruction. We explore the effect of $k$ in the AsynHDR system by comparing signal enhancement performance and ultimately chose $k=10$ based on the results, as shown in Fig.~\ref{fig:SNR curves}.
Additionally, other types of temporal weighting functions, such as quadratic or higher-order polynomial functions, can also be employed as temporal weighting strategies to enhance the signal. The denoising results of other weighting functions are measured and compared in our SNR test, as shown in Table~\ref{fig:SNR curves}. It can be observed that among numerous strategies, exponential weighting achieves state-of-the-art results. Therefore, we use the exponential function in this context and optimize its parameters.

\begin{table}[t]
\centering
\small
\caption{The SNR results for different temporal weighting methods, the h-poly term represents the SNR calculated using best high-order polynomial ($f(t)=(1-t)^5$) weighted result. }
\begin{tabular}{c|c|c}
      \hline
      \textbf{Weighting Method} & \textbf{c-map adjust} & \textbf{Mean SNR/(dB)}\\
      \hline
      raw integral & $\times$ & 14.98\\ 
      linear weighted& $\times$ & 15.99\\
      quadratic weighted& $\times$ & 16.32\\
      h-poly weighted& $\times$ & 16.56\\
     \textbf{Ours} & $\times$ & \textbf{16.61}\\
      \hline
      raw integral& $\checkmark$ & 21.74\\ 
      linear weighted& $\checkmark$ & 24.99\\
      quadratic weighted& $\checkmark$ & 26.43\\
      h-poly weighted& $\checkmark$ & 27.52\\
      \textbf{Ours} & $\checkmark$ & \textbf{27.67}\\
      \hline
    \end{tabular}
\end{table}

\begin{figure*}[htp]
  \centering
  \includegraphics[width=0.8\textwidth]{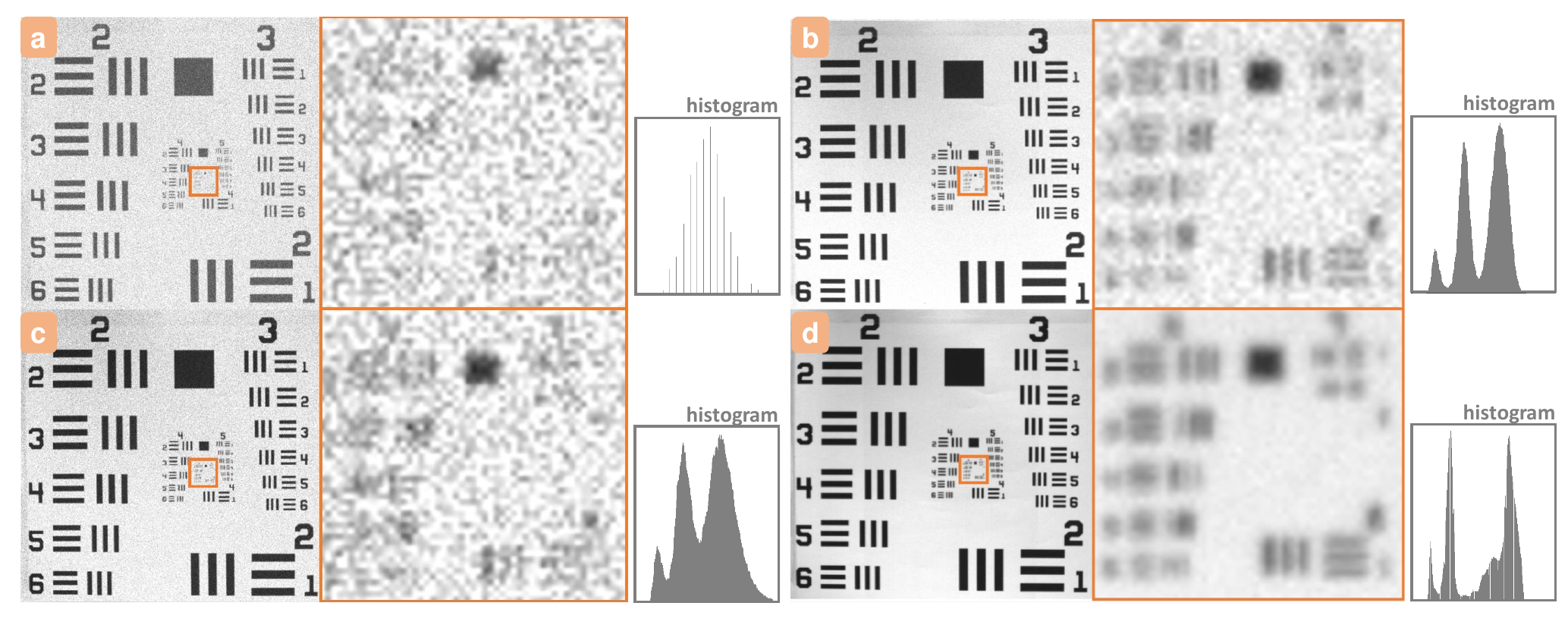}
  \caption{Imaging results of event flow using different reconstruction methods.For each algorithm, we provide the zoom-in images of the orange block.
  (a) Directly accumulate events to image (raw integral + c-map adjust). (b) Imaging by exponentially temporal weighted summation of events (ours + c-map adjust). (c) The linear temporal weighted summation of events to image (linear + c-map adjust). (d) Imaging results of the frame camera with identical resolution and sensor size under the same optical system (GT).
  }
  \label{fig:weighting result}
\end{figure*}
\begin{figure*}[htp]
  \centering
  \includegraphics[width=0.8\textwidth]{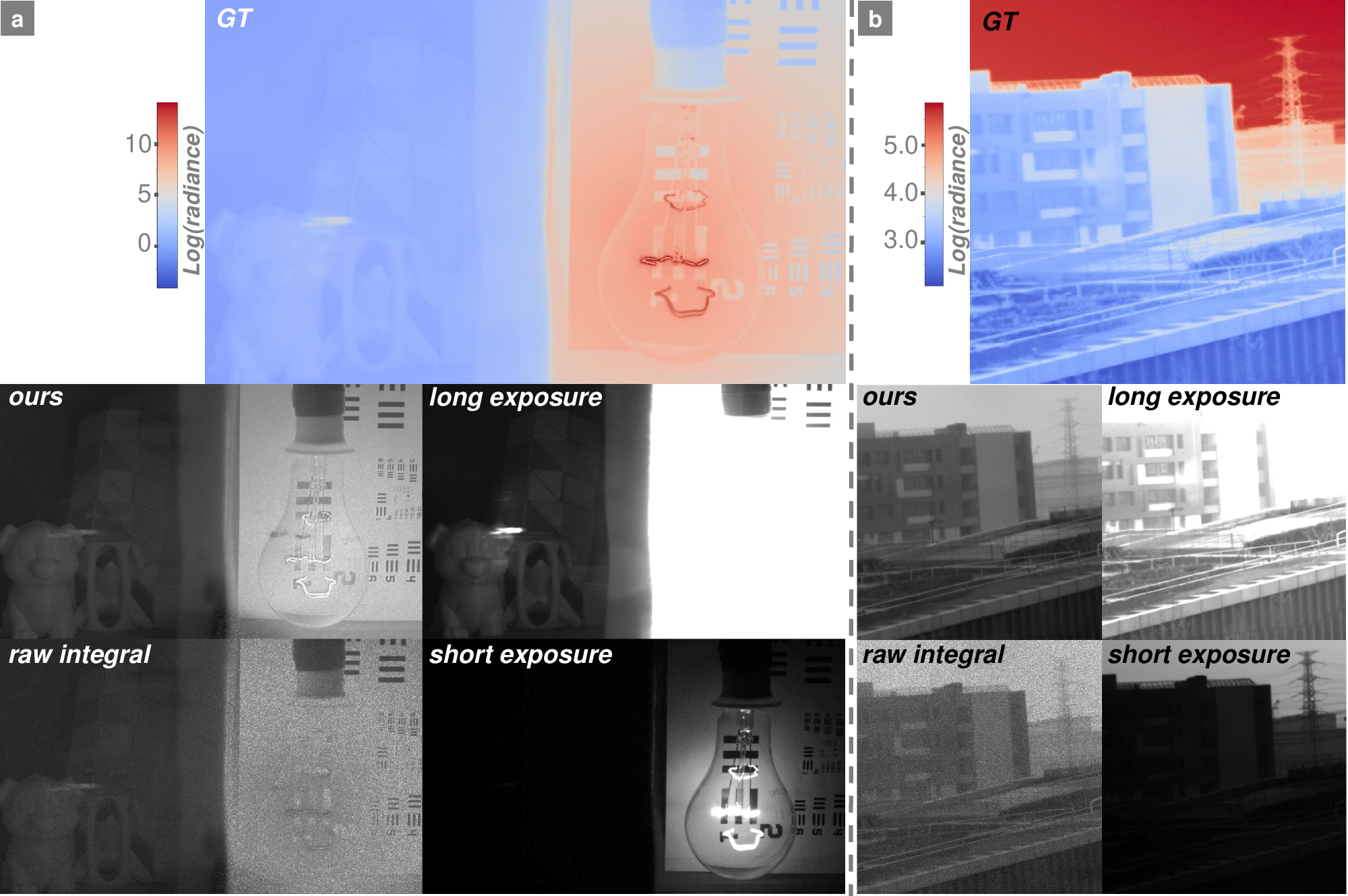}
  \caption{HDR imaging through the AsynHDR system. After carefully selecting different exposure times to capture 10 images using the frame camera in the system, the MEF method \cite{Debevec_Malik_2008} is employed to obtain the real scene radiance as a reference (GT). Our method is compared with frame-based cameras under long and short exposures to showcase HDR performance. Additionally, the directly integrating events (raw integral) method is also included for reference, to demonstrate the enhancement achieved by our algorithm.
  (a) The imaging result in scene with light source. (b)The imaging result under outdoor scenarios. 
  }
  \label{fig:real scenario test}
\end{figure*}
Additionally, we present the actual imaging results of different temporal weighting methods, referring to the histograms of the results in Fig.~\ref{fig:weighting result}. It can be observed that the image produced by our method in Fig.~\ref{fig:weighting result} (b) has more gray levels compared to raw integral in Fig.~\ref{fig:weighting result} (a), indicating better imaging quality.

We selected two challenging scenes for system real scene evaluation: The first scene involves simultaneous capturing of a dark box and an incandescent lamp to assess performance in extreme HDR scenarios (Fig.~\ref{fig:real scenario test}a). The second scene depicts an outdoor setting with a bright afternoon sky as the background (Fig.~\ref{fig:real scenario test}b), demonstrating the system's HDR performance in open outdoor environments. Considering the inherent limitation of active light triggered methods \cite{han2023high,takatani2021event} in imaging outdoor and light source included scenes, this set of experiments further validates the superiority of our system.

\section{Conclusion}
In this paper, we proposed an approach for constructing HDR imaging systems using asynchronous sensors, addressing HDR challenges through asynchronous sampling. 
Our experiments in HDR scenarios validate that the DVS can independently serve as a sensor to construct a multi-scene robust imaging system.
This implies, using the approach presented in this paper, we can replace frame-based cameras with DVS as the sensor for devices such as mobile phone and auto-pilot vehicles, rather than using it as an auxiliary for imaging. 

Although AsynHDR system effectively addressed HDR challenges, its frame rate is constrained to 20fps due to the bandwidth limitations of the DVS sensor, and it faces limitations in handling fast-moving scenes due to the scene radiance information's temporal coding. However, with advancements of DVS sensor, we anticipate future improvements in the system's frame rate, and plan to explore solutions for motion scenes in our future work. Moreover, using a DVS sensor designed with a Bayer matrix, AsynHDR can achieve color HDR imaging, similar to a frame-based RGB camera.

\begin{backmatter}
\bmsection{Disclosures}The authors declare that there are no conflicts of interest related to this article.
\bmsection{Data availability}Data underlying the results presented in this paper are not publicly available at this time but may be obtained from the authors upon reasonable request.
\end{backmatter}

\bibliography{main}

\end{document}